\documentclass{elsart}
\usepackage{graphicx,amsmath,amssymb, amsfonts}

\begin{document}

\begin{frontmatter}


  
\title{Comparison of time splitting and backsubstitution methods for
       integrating Vlasov's equation with magnetic fields}


\author{H. Schmitz} and
\author{R. Grauer}

\address{Theoretische Physik I, Ruhr-Universit\"at Bochum, 44780 Bochum,
Germany}

\begin{abstract}
  The standard approach for integrating the multidimensional Vlasov
  equation using grid based, conservative schemes is based on a time
  splitting approach. Here, we show that although the truncation error is
  of second order, time splitting can introduce systematic heating of the
  plasma. We introduce a backsubstitution method, which not only avoids
  this deficiency but also is computationally less expensive. The general
  approach is demonstrated in conjunction with Boris' scheme for
  evaluating the characteristics.
\end{abstract}

\end{frontmatter}

\section{Introduction}

Vlasov's equation is fundamental for numerous problems in plasma
theory. This kinetic equation describes the behaviour of the single
particle distribution functions of a collisionless plasma under the
influence of electric and magnetic fields. Coupled with the equations
for the electromagnetic fields and the evaluation of the moments of
the distribution functions one obtains a highly nonlinear system of
differential and integral equations. Only a few very simple problems
can be solved analytically. For this reason numerical simulations of
Vlasov's equation have become an important tool for theoretical plasma
physics.

One type of computer simulation approach integrates the distribution
function directly on a high--dimensional numerical grid in phase space.
Here one dimension is needed for every space component and for every
velocity component. Following the original work by Cheng and Knorr
\cite{CHNG76} much progress has been made on improving the accuracy of the
advection schemes.

The integration of the distribution function can be carried out in a
number of different ways: The simplest schemes are finite difference
schemes. They are relatively easy to implement but suffer from numerical
instabilities and nonpositivity of the distribution function.
Conservation laws such as the conservation of particle number can be
implemented but complicate the scheme greatly \cite{ARA66}. Conservative
methods, on the other hand, discretise the distribution function by
integrating over the numerical grid cells \cite{FIJ99}. The advantage of
these methods lies in the fact that the particle number is naturally
conserved and no artificial sources or sinks of particles are introduced.
Semi-Lagrangian methods (e.g. \cite{SON98}) follow the characteristics
backwards and interpolate the distribution function at the origin of the
characteristic. The interpolated value is then transported forward to the
grid points. Semi-Lagrangian methods do not naturally conserve the particle
number but can easily be made to preserve positivity.

Most of the above methods are, however, developed for a one--dimensional
advection problem. When used for a one--dimensional electrostatic system
in which the physical phase space is two--dimensional, a time splitting
method is employed which was already proposed in \cite{CHNG76}. Although
Semi-Lagrangian methods in principle allow to integrate the distribution
function directly on the high--dimensional grid, the time splitting
technique is also used to simplify the computation \cite{FIL03}. The
general idea is that in higher dimensions this time splitting can be
generalised in a straightforward way \cite{CAL01}.

We will show in this paper that, when including a magnetic field, this
simple time splitting --- although second order --- can cause dissipation
due to errors which are always in the same direction. This implies that the
temperature of the system will increase systematically. We will also
present an alternative method which we named {\em backsubstitution }
method. The backsubstitution method not only eliminates the problems of the
time splitting method but is also computationally less expensive.

In section 2 we will present the basic underlying equations. Section 3
will describe the time splitting method and show how a systematic error
develops. In section 4 we will present the backsubstitution method  which
we will apply to Boris' scheme in section 5. Section 6 discusses
simulations of Bernstein waves using the different schemes to provide a
comparison. Section 7 gives some concluding remarks.

\section{General Problem}

The basis of the kinetic plasma description is the distribution
function $f({\bf x}, {\bf v},t)$ which expresses the particle density
in phase space. Here $f(\mathbf{x}, \mathbf{v},t)\;d^3x\;d^3v$ is
the number of particles in a phase space volume $d^3x\;d^3v$ located
at $(\mathbf{x}, \mathbf{v})$ at time $t$. In a collisionless plasma
the evolution of the distribution function is given by Vlasov's
equation
\begin{equation}
\frac{\partial f}{\partial t} 
        + {\bf v}\cdot \nabla_{\mathbf{x}} f
        + \frac{q}{m} \left( 
                \mathbf{E} + \mathbf{v} \times \mathbf{B}
        \right) \cdot \nabla_{\mathbf{v}} f
= 0,\label{VlasovOrig}
\end{equation}
where $\mathbf{E}$ and $\mathbf{B}$ are the electric and magnetic fields
which have to be determined self-consistently. Vlasov's equation describes
the advection of values of the distribution function along particle
characteristics given by Newton's law of motion.

One central property of Vlasov's equation is the conservation of the phase
space density, which directly translates into a conservation of mass and
charge in a closed system. For this reason it is natural to use a
conservative scheme for simulating Vlasov's equation (for 1--dimensional
schemes, see e.g. \cite{FIL01}). Today, a diversity of Eulerian schemes,
all with high accuracy and different advantages and disadvantages, are
available (see e.g. \cite{FIL03,ARB02} and references therein). These
schemes normally solve the one--dimensional advection problem,
\begin{equation}
\partial_t f(x,t) + u \partial_x f(x,t) = 0. \label{Transport1d}
\end{equation}
By integrating over a finite time step one obtains
\begin{equation}
\int\limits_{x_{i-1/2}}^{x_{i+1/2}} f(x,t^{n+1}) \; dx
= \int\limits_{X(t^n, t^{n+1}, x_{i-1/2})}^{X(t^n, t^{n+1},x_{i+1/2})}
f(x,t^n) \; dx. \label{CharacTransportInt}
\end{equation}
Here $X(s,t,\xi)$ denotes the characteristic with parameter $s$ that
satisfies \mbox{$X(t,t,\xi) = \xi$}.

For the one dimensional, electrostatic Vlasov--Problem 
\begin{equation}
\frac{\partial f}{\partial t} 
        + v \partial_x f
        + \frac{q}{m} E \partial_v f
= 0,\label{VlasovOrig1d}
\end{equation}
a splitting technique is then usually employed. Here one integrates the
advection in the $x$--direction by $\Delta t/2$, then in $v$--direction by
$\Delta t$ and then again in $x$--direction by $\Delta t/2$. This produces
a second order scheme which can be written as
\begin{equation}
T_x(\Delta t/2) T_v(\Delta t) T_x(\Delta t/2).
\end{equation}
Here $T_k$ denotes the numeric advection operator in the $k$--dimension.

\section{Time splitting\label{SecTimeSplit}}

The success of the time splitting for the one dimensional electrostatic
problem motivates a common suggestion to extend the splitting technique to
treat higher--dimensional systems. Since the spatial dimensions are
completely independent of each other this results in the following second
order scheme for the full three dimensional system 
\begin{multline}
T_x(\Delta t/2) T_y(\Delta t/2) T_z(\Delta t/2)\\
T_{v_x}(\Delta t/4) T_{v_y}(\Delta t/2) T_{v_x}(\Delta t/4)
T_{v_z}(\Delta t)
T_{v_x}(\Delta t/4) T_{v_y}(\Delta t/2) T_{v_x}(\Delta t/4)\\
T_x(\Delta t/2) T_y(\Delta t/2) T_z(\Delta t/2).
\end{multline}

In each of these sub-steps a one dimensional transport equation of type
(\ref{Transport1d}) is solved.
For each of these equations the characteristics are calculated and then
projected onto the corresponding direction. This implies that even for a
hypothetical {\em exact} one--dimensional integration scheme the
characteristics are still only approximated by a second order time
splitting scheme. For the following discussion, we will consider only the
velocity part of the integration scheme since this determines how well the
particle temperatures are described.

For a purely electrostatic system the above integration behaves well
and errors only occur due to repeated application of the advection
scheme. The reason for this is the independence of the change of the
velocity component $\Delta v_k$ on the velocity $v_k$. With a magnetic
field, however, the change of velocity $\Delta v_k$ over a finite time
step does depend on the velocity $v_k$. Here we assume that the
integration scheme for the characteristics is at least second order.
To investigate the error caused by this method, we take the exact
characteristic in $\mathbf{v}$--space and approximate it using the
time splitting scheme.  During integration of the characteristic, the
electromagnetic fields are assumed to be constant. Without loss of
generality we let $\mathbf{B} = B \mathbf{\hat{z}}$ and move the
origin in velocity space to $\mathbf{v}_0 =
\mathbf{E}\times\mathbf{B}$. For simplicity we assume $E_z=0$.
$E_z\ne0$ would only add a constant acceleration in the
$v_z$--direction, and leads to the same result.

In this setup the characteristics in velocity space are simple concentric
circles around the origin and we can neglect the $v_z$ coordinate
completely.  During the time interval $\Delta t$ the whole
$v_x$,$v_y$--plane rotates by an angle $\phi = \Delta t qB/m$. Taking
a velocity
\begin{equation}
\mathbf{v} = (v_x, v_y)
\end{equation}
we split the rotation into three steps according to the time--splitting 
scheme
\begin{equation}
T_{v_x}(\Delta t/2) T_{v_y}(\Delta t) T_{v_x}(\Delta t/2).
\label{TimeSplitting}
\end{equation}
This results in the following
\begin{alignat}{1}
\mathbf{v}^a &= (v_x \cos(\phi/2) - v_y \sin(\phi/2), v_y) \label{SchemeA1} \; ,\\
\mathbf{v}^b &= (v^a_x , v^a_y \cos(\phi) + v^a_x \sin(\phi)) \; ,
\label{SchemeA2}\\
\mathbf{v}^{\text{new}} &= (v^b_x \cos(\phi/2) - v^b_y \sin(\phi/2), v^b_y) \; . 
\label{SchemeA3}
\end{alignat}
Inserting $\mathbf{v}^a$ into $\mathbf{v}^b$ and then $\mathbf{v}^b$ into 
$\mathbf{v}^{\text{new}}$ results in a lengthy expression for 
$\mathbf{v}^{\text{new}}$. Taking the norm of $\mathbf{v}^{\text{new}}$ and
expanding this expression for small angles
$\phi$, i.e. small time steps $\Delta t$, gives
\begin{equation}
\left(  {v^{\text{new}}}\right)^2 = v_x^2 + v_y^2 - \left( v_x^2+\frac{v_y^2}{2} \right)\phi^2
+ \mathcal{O}(\phi^3). \label{TimeSplitSecondOrder}
\end{equation}
By construction this is, of course, second order in $\phi$. However, one can
see that the second order error is {\em always} negative and thus
introduces a systematic error. 

The time splitting method (\ref{TimeSplitting}) can also be interpreted as
performing the individual steps (\ref{SchemeA1})--(\ref{SchemeA3}) in first
order in $\phi$. This corresponds to
\begin{alignat}{1}
\mathbf{v}^a &= (v_x  - v_y \phi/2, v_y) \label{SchemeB1} \; ,\\
\mathbf{v}^b &= (v^a_x , v^a_y + v^a_x \phi) \; ,
\label{SchemeB2}\\
\mathbf{v}^{\text{new}} &= (v^b_x  - v^b_y \phi/2, v^b_y) \; .
\label{SchemeB3}
\end{alignat}
With respect to equation (\ref{Transport1d}) this scheme is obtained by holding
$u$ constant for each step. Taking the square of this
$\mathbf{v}^{\text{new}}$ results in
\begin{equation}
{v^{\text{new}}}^2 = v_x^2 + v_y^2 + \frac{\phi^3}{2}v_x v_y +
\mathcal{O}(\phi^4).
\end{equation}
One can observe that in this case the second order disappears, and the
third order error is not systematic, but depends on the signs of $v_x$ and
$v_y$. In the following we will refer to eqs
(\ref{SchemeA1})--(\ref{SchemeA3}) as {\em scheme A} and
(\ref{SchemeB1})--(\ref{SchemeB3}) as {\em scheme B}.

\begin{figure}
\begin{center}
\includegraphics[width=11cm]{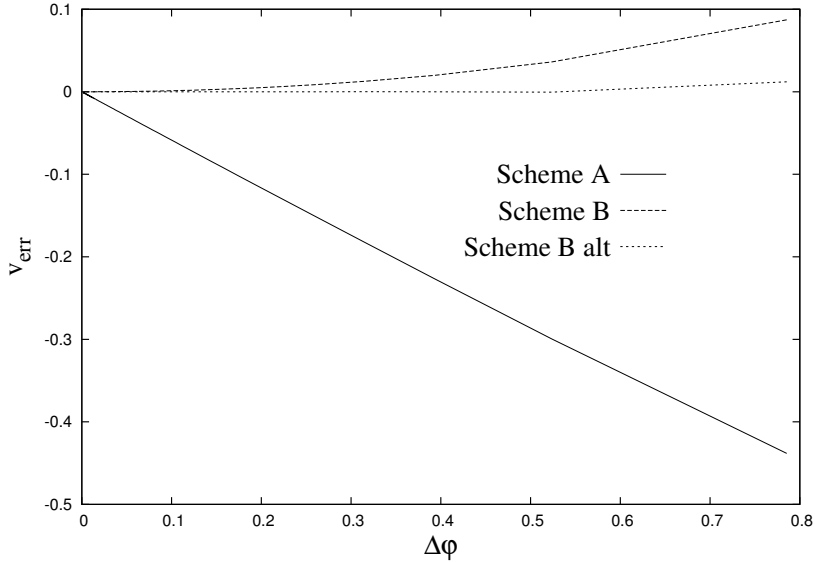}
\end{center}
\caption{Error $v_{err}$ of the magnitude of the velocity after a quarter
gyration depending on the angle of an individual step. Curves are plotted
for the schemes A, B and scheme B with alternating order of the
$v_x$--$v_y$ integration.
\label{FigVelErrFull}}
\end{figure}

In Fig. \ref{FigVelErrFull} the error of the magnitude $|\mathbf{v}|$ of
the velocity after a quarter gyration is plotted against the rotation
angle $\Delta \varphi$ of the individual step. The error is normalised to
the initial velocity. The solid line shows the result of scheme A while
the dashed line shows the result of scheme B.  The dotted line represents
the result of scheme B with alternating order of the $v_x$--$v_y$
integration. For scheme A, a total error of about 2.5\% is accumulated
after a quarter gyration when $\Delta \varphi \approx 0.045$. After a full
gyration the error sums up to 10\% (not shown). This value of $\Delta
\varphi$ corresponds to roughly 140 integration steps for the full circle.
Using less steps, i.e. larger $\Delta \varphi$ results in even larger
errors.

To understand the direction of the error we note, that in eq.
(\ref{CharacTransportInt}) the characteristics are integrated
backwards from a grid point to obtain the source of the distribution
function for that grid point. The distribution function is then
transported from that source to the grid in some manner that depends
on the numerical scheme. The negative sign in the second order of eq.
(\ref{TimeSplitSecondOrder}) implies that the source is always located
closer to the rotation centre than the grid point. Thus the values of
the distribution function are transported outwards from the rotation
centre. This results in an effective heating of the distribution
function.

\begin{figure}
\begin{center}
\includegraphics[width=11cm]{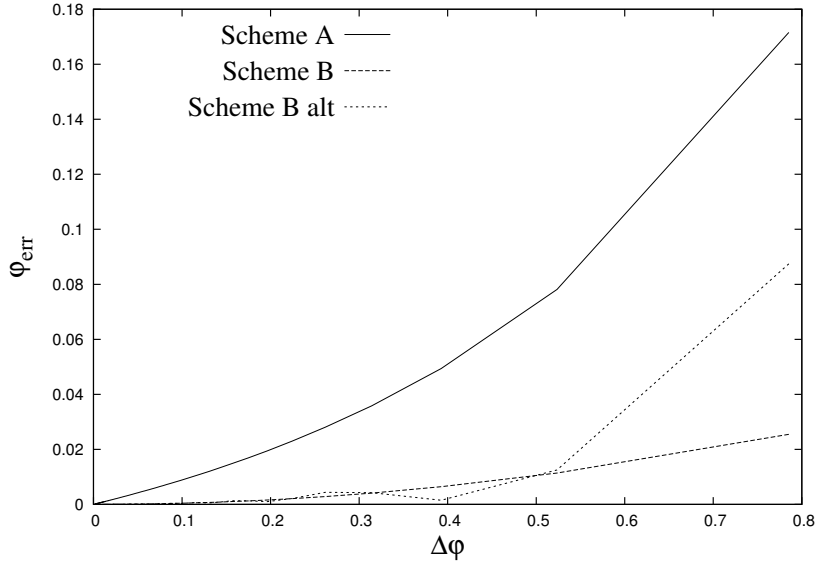}
\end{center}
\caption{Error $\phi_{err}$ of the phase of the velocity after a quarter
gyration depending on the angle of an individual step. Curves are plotted
for the schemes A, B and scheme B with alternating order of the
$v_x$--$v_y$ integration.
\label{FigPhaseErrFull}}
\end{figure}

Using scheme B the errors are smaller but not zero. Here an error of 2\%
is observed when $\Delta \varphi \approx 0.4$. This is equivalent to
roughly 16 steps for a full gyration. Because the direction of the error
in scheme B depends on the values of $v_x$ and $v_y$, one can further
increase the accuracy by alternating the order of the splitting. In the two
dimensional case considered here this simply implies alternating the roles
of $v_x$ and $v_y$. Using the alternating scheme the overall error in the
velocity magnitude is reduced to almost zero. However, when looking at the
relative phase error after a quarter gyration (Fig. \ref{FigPhaseErrFull})
no significant improvement can be observed. While scheme A still shows the
largest error, the errors for scheme B with and without alternating
oder of integration are roughly comparable  up to a $\Delta \varphi$ of
0.5. For this value of $\Delta \varphi$ the phase error is approximately
1\%. 

When the Vlasov equation is solved on a discretised grid errors are
worse but the main sources of these errors are highlighted by the above
analytical argument. 

\section{Backsubstitution}

In this section we want to present an alternative method for integrating
Vlasov's equation that does not suffer from the above drawbacks. Here we
will present first the general idea of this {\em backsubstitution method}
and then write down the equations for the general system described above.

Suppose we are given a one dimensional integration scheme for the
transport eq. (\ref{Transport1d}). To create a scheme for the
integration of the three--dimensional velocity space there is no other
choice but to split the full three--dimensional problem into a number
of one--dimensional substeps. For each of these substeps the
characteristics will be calculated and then projected onto the
direction of the advection step. We still have the freedom, which
characteristics to integrate and in which order to integrate them.

\begin{figure}
\begin{center}
\includegraphics[width=10cm]{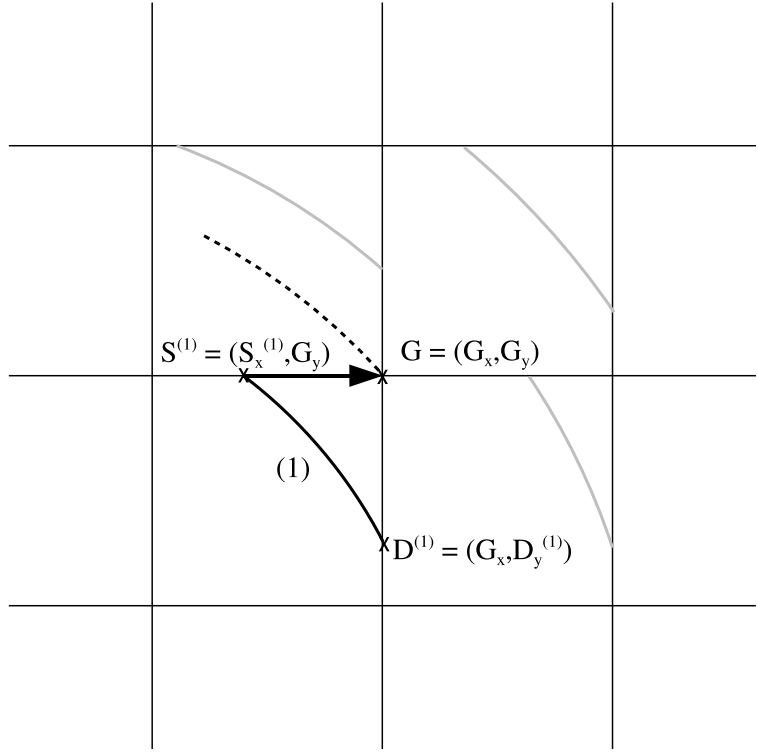}
\end{center}
\caption{Schematic diagram of the first step of integration of the 
characteristics using the backsubstitution algorithm. The distribution
function is shifted in the $v_x$--direction from the source point $S^{(1)}$
of the characteristic to the grid point $G$. Gray lines show corresponding
characteristics in the other cells. The dashed line is the characteristic
ending on $G$ which is important in the second step.
\label{FigSchematicBackSubs1}}
\end{figure}

\begin{figure}
\begin{center}
\includegraphics[width=10cm]{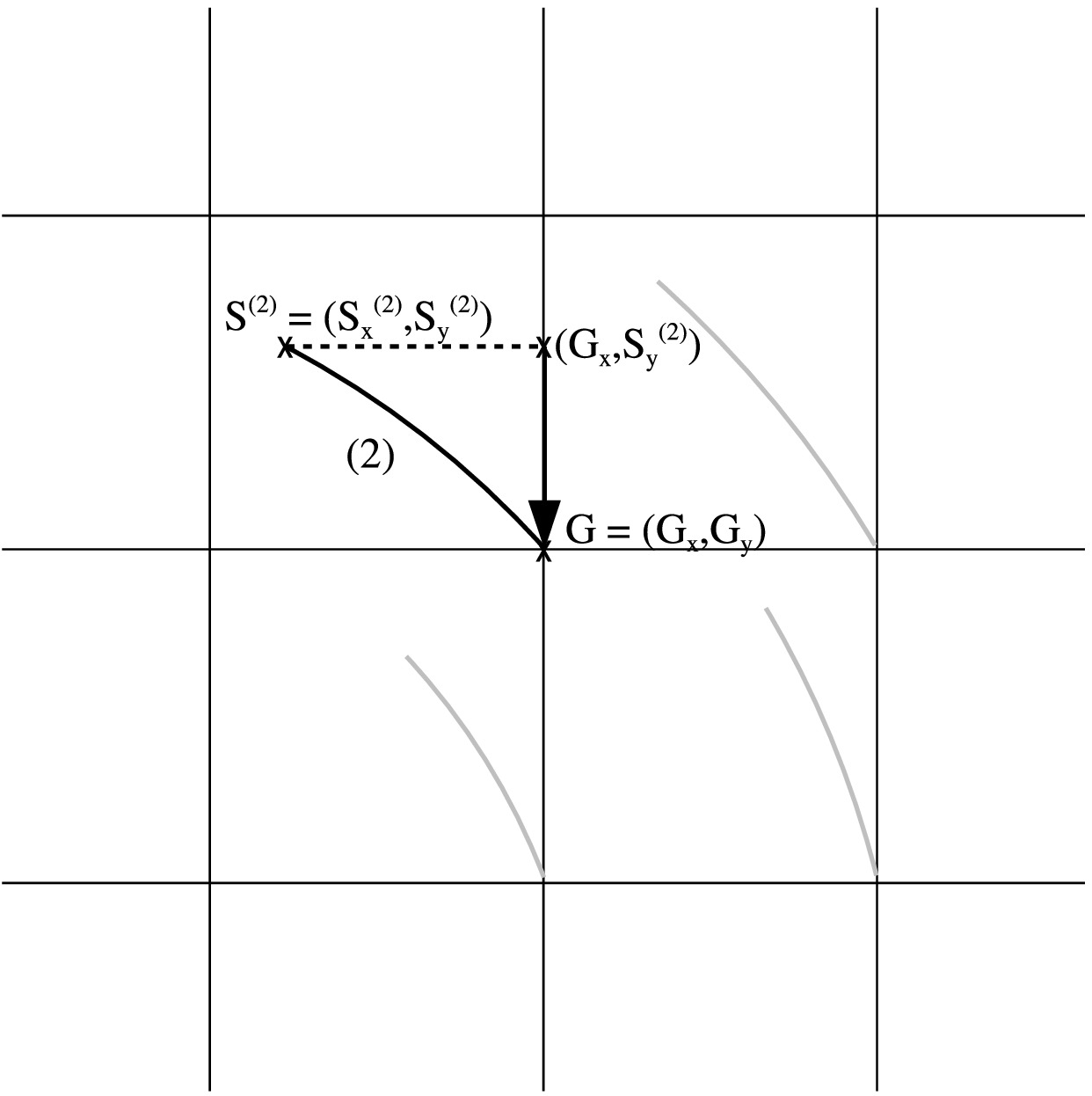}
\end{center}
\caption{Schematic diagram of the second step of integration of the 
characteristics using the backsubstitution algorithm. The distribution
function is shifted in the $v_y$--direction from the intermediate point 
to the destination point $G$ of the characteristic. Gray lines show 
corresponding characteristics in the other cells.
\label{FigSchematicBackSubs2}}
\end{figure}

To start with, let us again consider the standard case described in
the last section. Our aim is to formulate a splitting scheme in which
the characteristics are integrated exactly, and which uses the minimum
number of integration steps.  Since we can ignore the
$v_z$--direction, this means we want only two integration steps, one
for $v_x$, and one for $v_y$.

The distribution function is first shifted in the $v_x$, and then in the
$v_y$ direction. Figure \ref{FigSchematicBackSubs1} illustrates the first
step while Figure \ref{FigSchematicBackSubs2} illustrates the second step.
Both shifts together should transport the value of the distribution
function from a source point $S=(S_x, S_y)$ of a characteristic to its 
destination point $D=(D_x,D_y)$ with $S = X(t-\Delta t, t, D)$. Here the
indices $x$, $y$, and $z$ are used to denote the velocity components
$v_x$, $v_y$ and $v_z$. This means  that we aim to find a scheme such
that
\begin{equation}
f^{\text{new}}(D_x, D_y) = f^{\text{old}}(S_x, S_y).
\end{equation}

In the first step the shift in $v_x$ has to transport $f$ from $S$ to
an intermediate point $(D_x, S_y)$. In the semi-Lagrangian schemes which we are 
considering here, the characteristics are integrated backward from the grid
points. This implies that in the first step (1) the grid point $G$ has to
coincide with the intermediate point  $G=(D^{(1)}_x,S^{(1)}_y)$, or
equivalently $S^{(1)} = (S^{(1)}_x, G_y)$ and $D^{(1)} = (G_x, D^{(1)}_y)$.
We have displayed these characteristics in Figure
\ref{FigSchematicBackSubs1}. In this way the distribution function has
been shifted along $v_x$ according to 
\begin{equation}
f^{\text{inter}}(G_x, G_y) = f^{\text{old}}(S^{(1)}_x, G_y).
\end{equation}

Given a sufficiently smooth behaviour of the characteristics we can assume
that the interpolation scheme causes all other points of the distribution
function to be shifted accordingly. This is particularly true for the
characteristic that ends in the grid point $G$ (dashed line in Figure
\ref{FigSchematicBackSubs1}). This characteristic will be important in the
following step.

In the second step (Figure \ref{FigSchematicBackSubs2}) we, therefore, need
to choose the characteristic that ends in $G$. Then the source point
$S^{(2)}$ is given by $S^{(2)} = X(t-\Delta t, t, G)$.  The shift is
performed in the $v_y$ direction so that
\begin{equation}
f^{\text{new}}(G_x, G_y) = f^{\text{inter}}(G_x, S^{(2)}_y).
\end{equation}
Since in the first step we had (assuming again correct interpolation)
\begin{equation}
f^{\text{inter}}(G_x, S^{(2)}_y) = f^{\text{old}}(S^{(2)}_x, S^{(2)}_y),
\end{equation}
we finally have
\begin{equation}
f^{\text{new}}(G_x, G_y) = f^{\text{old}}(S^{(2)}_x, S^{(2)}_y).
\end{equation}

We now use this motivation to write down a general scheme for
three--dimensional velocity space. For every grid point $G$ we perform
the integration in three one--dimensional substeps, one for each
component $v_x, v_y, v_z$. For each integration a source coordinate
$S^{(1)}_x, S^{(2)}_y$ and $S^{(3)}_z$ is calculated from a
characteristic which does not necessarily pass through $G$. To find
$S^{(1)}_x$ for the $v_x$--integration we demand
\begin{alignat}{1}
G_x &= D^{(1)}_x\label{EqSystemX1} \; ,\\
G_y &= \left( X(t-\Delta t, t, D^{(1)}) \right)_y\label{EqSystemX2} \; ,\\
G_z &= \left( X(t-\Delta t, t, D^{(1)}) \right)_z \; . \label{EqSystemX3}
\end{alignat}
In general this is a nonlinear system of equations for the components
$D^{(1)}_y$ and $D^{(1)}_z$. The details of this system depend on the way the 
characteristics are calculated. Given $D^{(1)}$ one then has
\begin{equation}
S^{(1)}_x = \left( X(t-\Delta t, t, D^{(1)}) \right)_x \; ,
\end{equation}
and the integration can be performed from $S^1_x$ to $G_x$ in the
$v_x$--direction.

Similarly we demand for the $v_y$--integration
\begin{alignat}{1}
G_x &= D^{(2)}_x \label{EqSystemY1} \; ,\\
G_y &= D^{(2)}_y \label{EqSystemY2} \; ,\\
G_z &= \left( X(t-\Delta t, t, D^{(2)}) \right)_z \; . \label{EqSystemY3}
\end{alignat}
$D^{(2)}$ differs from $G$ only in the $v_z$--component. Once $D^{(2)}_z$ is found
we have
\begin{equation}
S^{(2)}_y = \left( X(t-\Delta t, t, D^{(2)}) \right)_y.
\end{equation}
Again the integration is now performed from $S^{(2)}_y$ to $G_y$ in the
$v_y$--direction.

The $v_z$--integration finally is straightforward. Since $D^{(3)} = G$ we have
\begin{equation}
S^{(3)}_z = \left( X(t-\Delta t, t, G) \right)_y
\end{equation}
and the integration is performed from $S^{(3)}_z$ to $G_z$ in the
$v_z$--direction.

We want to emphasise that this scheme integrates the characteristics
exactly, which means that in terms of Figs. \ref{FigVelErrFull} and
\ref{FigPhaseErrFull} the backsubstitution scheme has errors which are
{\em exactly zero}.

\section{Application to Boris scheme}

The main task now is to calculate the characteristics or their approximations
in the presence of a magnetic field. A commonly used approach is
the Boris scheme \cite{BIR85}. Here the integration step is formulated as 
an implicit finite difference scheme
\begin{equation}
\frac{\mathbf{v}^{n+1} - \mathbf{v}^n}{\Delta t} = \frac{q}{m}
\left( 
  \mathbf{E} + \frac{\mathbf{v}^{n+1} + \mathbf{v}^n}{2} 
  \times \mathbf{B}
\right)
\end{equation}
The electric and magnetic forces are separated,
\begin{alignat}{1}
\mathbf{v}^- &= \mathbf{v}^n + \frac{\Delta t}{2}\frac{q}{m}\mathbf{E} \; ,
\label{EqDefVPlus}\\
\mathbf{v}^+ &= \mathbf{v}^{n+1} - \frac{\Delta t}{2}\frac{q}{m}\mathbf{E} \; ,
\label{EqDefVMinus}
\end{alignat}
leading to
\begin{equation}
\frac{\mathbf{v}^+ - \mathbf{v}^-}{\Delta t} = \frac{q}{2m}
\left( \mathbf{v}^+ + \mathbf{v}^- \right) \times \mathbf{B} \; .
\end{equation}
The transformation from $\mathbf{v}^-$ to $\mathbf{v}^+$ is a pure
rotation with an angle $\theta$ where
\begin{equation}
\left| \tan \frac{\theta}{2}\right| = \frac{\Delta t}{2}\frac{qB}{m}.
\end{equation}
For small angles $\theta$ this is close to the exact angle
$\theta_{\text{exact}} = \Delta t q B /m$.

In contrast to the original scheme of Boris, we aim to trace the
characteristics backward in time. This means we want to find $\mathbf{v}^-$
in terms of $\mathbf{v}^+$. We thus reverse the original scheme and rotate 
$\mathbf{v}^+$ by $\vartheta$.
To implement this rotation the vectors $\mathbf{t}$ and $\mathbf{s}$ are
defined
\begin{equation}
\mathbf{t} = \frac{-\Delta t}{2}\frac{q \mathbf{B}}{m} \;\; , \;\;\;
\mathbf{s} = \frac{2\mathbf{t}}{1+t^2}. \label{EqDefTAndS}
\end{equation}
Then the rotation is performed in two steps
\begin{equation}
\mathbf{v}' = \mathbf{v}^+ + \mathbf{v}^+ \times \mathbf{t} \; ,
\label{EqDefVPrime}
\end{equation}
and
\begin{equation}
\mathbf{v}^- = \mathbf{v}^+ + \mathbf{v}' \times \mathbf{s} \; .
\label{EqCalcVMinus}
\end{equation}
This scheme now supplies $S = \mathbf{v}^n = (v_x^n, v_y^n, v_z^n)$ in 
terms of $D = \mathbf{v}^{n+1} = (v_x^{n+1}, v_y^{n+1}, v_z^{n+1})$. 
To facilitate the further calculations we insert (\ref{EqDefVPrime}) into
(\ref{EqCalcVMinus}) and separate $\mathbf{v}^-$ into it's components
\begin{alignat}{1}
v^-_x &= (1 - s_y t_y - s_z t_z) v^+_x + (s_y t_x + s_z) v^+_y 
  + ( s_z t_x - s_y) v^+_z \; , \label{EqCalcVMinusX}\\
v^-_y &= ( s_x t_y-s_z) v^+_x + (1 - s_x t_x - s_z t_z) v^+_y 
  + ( s_z t_y + s_x) v^+_z \; , \label{EqCalcVMinusY}\\
v^-_z &= (s_x t_z + s_y) v^+_x + (s_y t_z - s_x) v^+_y 
  + (1 - s_x t_x - s_y t_y) v^+_z \; . \label{EqCalcVMinusZ}
\end{alignat}

We now need to solve the systems of equations (\ref{EqSystemX1} --
\ref{EqSystemX3}) and (\ref{EqSystemY1} -- \ref{EqSystemY3}) for the first
and the second backsubstitution step. As stated before, the third step is 
straightforward since $D^{(3)}$ is already known.
The complete problem can be written in the form 
\begin{alignat}{1}
v_x^n &= v_x^n(v_x^{n+1}, v_y^n, v_z^n) \; ,\\
v_y^n &= v_y^n(v_x^{n+1}, v_y^{n+1}, v_z^n)\; ,\\
v_z^n &= v_z^n(v_x^{n+1}, v_y^{n+1}, v_z^{n+1}) \; .
\end{alignat}
Since the bijections between $v^n$ and $v^-$, on one hand, and $v^{n+1}$ and
$v^+$, on the other hand, are trivial (see eqs (\ref{EqDefVPlus}) and
(\ref{EqDefVMinus})) it is sufficient to formulate the three steps
\begin{alignat}{1}
v_x^- &= v_x^-(v_x^+, v_y^-, v_z^-) \; , \label{EqBorisBackX}\\
v_y^- &= v_y^-(v_x^+, v_y^+, v_z^-) \; , \label{EqBorisBackY}\\
v_z^- &= v_z^-(v_x^+, v_y^+, v_z^+) \; . \label{EqBorisBackZ}
\end{alignat}

To find (\ref{EqBorisBackX}) we take eqs (\ref{EqCalcVMinusY}) and
(\ref{EqCalcVMinusZ}) and solve for $v^+_y$ and $v^+_z$ giving
\begin{alignat}{1}
v^+_y &= \frac{(1-s_x t_x-s_y t_y)v^-_y - (s_x + s_z t_y)v^-_z
  + (s_x t_y (m - 1) - s_x n_y + s_x s_y + s_z )v^+_x}
  {m (s_x t_x - 1) + 1 + s_x (s_x - t_x - n_x)} \\
v^+_z &= \frac{(s_x - s_y t_z)v^-_y + (1-s_x t_x-s_z t_z)v^-_z
  + (s_x t_z(m - 1) - s_x n_z - s_y + s_x s_z)v^+_x}
  {m (s_x t_x - 1) + 1 + s_x (s_x - t_x - n_x)}
\end{alignat}
where $m = \mathbf{s}\cdot\mathbf{t}$ and $\mathbf{n} =
\mathbf{s}\times\mathbf{t}$. These can be inserted into
(\ref{EqCalcVMinusX}) which then provides the expression
(\ref{EqBorisBackX}) for the first step.

For the second step (\ref{EqBorisBackY}) only eq (\ref{EqCalcVMinusZ}) has
to be solved for $v^+_z$ giving
\begin{equation}
v^+_z = \frac{ - (s_y + s_x t_z) v^+_x + 
  (s_x - s_y t_z) v^+_y + v^-_z}{1 - s_x t_x - s_y t_y}.
\end{equation}
With this, $v^+_z$ can be substituted in eq (\ref{EqCalcVMinusY}) giving
$v^-_y$ in the form (\ref{EqBorisBackY}).

By virtue of equation (\ref{EqCalcVMinusZ}) the $z$--component  is already
given in the form (\ref{EqBorisBackZ}). Thus, no further calculation has
to be done for the third step.

Finally we want to discuss the error of Boris' scheme combined with the
backsubstitution method. We again investigate the same problem as
formulated in section \ref{SecTimeSplit} where the velocity vector is
rotated around the origin. While Boris' scheme introduces a phase error in
this rotation, the magnitude of the velocity is preserved. Using this
combined scheme in a grid based Vlasov solver implies that the only diffusion
in the system originates from the reconstruction of the distribution
function. 

\section{Bernstein Waves}

We have applied the schemes described above to the simulation of Bernstein
waves in a periodic system. These are electrostatic waves propagating at a
right angle to a given constant magnetic field \cite{NICH83,BER58}. The
ions are treated as a static neutralising background, while the electrons
oscillate in the electrostatic field. We assume that $\mathbf{B} = B_z
\mathbf{\hat{z}}$ and the wavevector $\mathbf{k} = k_x \mathbf{\hat{x}}$. Then the
dispersion relation can be written as
\begin{equation}
1 + \frac{2 \omega^2_{pe}}{\sqrt{\pi} \Omega^2}
\sum\limits_{n=0}^{\infty} A_n(w) k^{2n} 
\frac{M\left(n+3/2, 2 n+2, -k^2\right)\Gamma\left(n+1.5\right)}
{\Gamma\left(2n+2\right)} = 0 \; ,
\end{equation}
with
\begin{equation}
A_n(w) = \frac{\left(2n+1\right)\left(n^2+n-w^2\right)}
{\left(n^2-w^2\right)\left(\left(n+1\right)^2-w^2\right)}.
\end{equation}
Here we used $w = \omega / \Omega$, $k = v_{th} k_x /\Omega$, $\Omega = e
B_z / m_e$ is the electron cyclotron frequency and $\omega_{pe}$ is the 
electron plasma frequency. $m_e$ is the electron mass and $e$ is the
electron charge. $\Gamma$ is the gamma function and $M$ is Kummer's
confluent hypergeometric function. We chose $\omega^2_{pe} = \Omega^2$ for
all simulations. For a given $k_x$, the above dispersion relation has an
infinite number of solutions for $\omega$. We performed the simulations in
one space and three velocity dimensions, $(x, v_x, v_y, v_z)$. Although
two velocity dimensions would be sufficient for this system, we keep the
$v_z$--dimension to make the results transferable to electromagnetic
simulations in which the magnetic field is not fixed.  The simulation box
has a length $L$ which was resolved with 64 grid cells. The velocity space
was sampled with 50 grid cells in each direction in the interval from
$-4v_{th}$ to $+4v_{th}$ . The length of the box is chosen to $L =
2\pi/k_x$ so that exactly one wavelength of the Bernstein mode fits into
the system. In this way the size of a grid cell in space is $\Delta x =
2\pi/ 64 k_x$. The timestep was chosen such that the CFL--condition is
satisfied $\Delta t = \Delta x / 5 v_{th}$. For the integration of the
distribution function on the grid we use a flux conservative and positive
scheme \cite{FIL01}.

\begin{figure}
\begin{center}
\includegraphics[width=10cm]{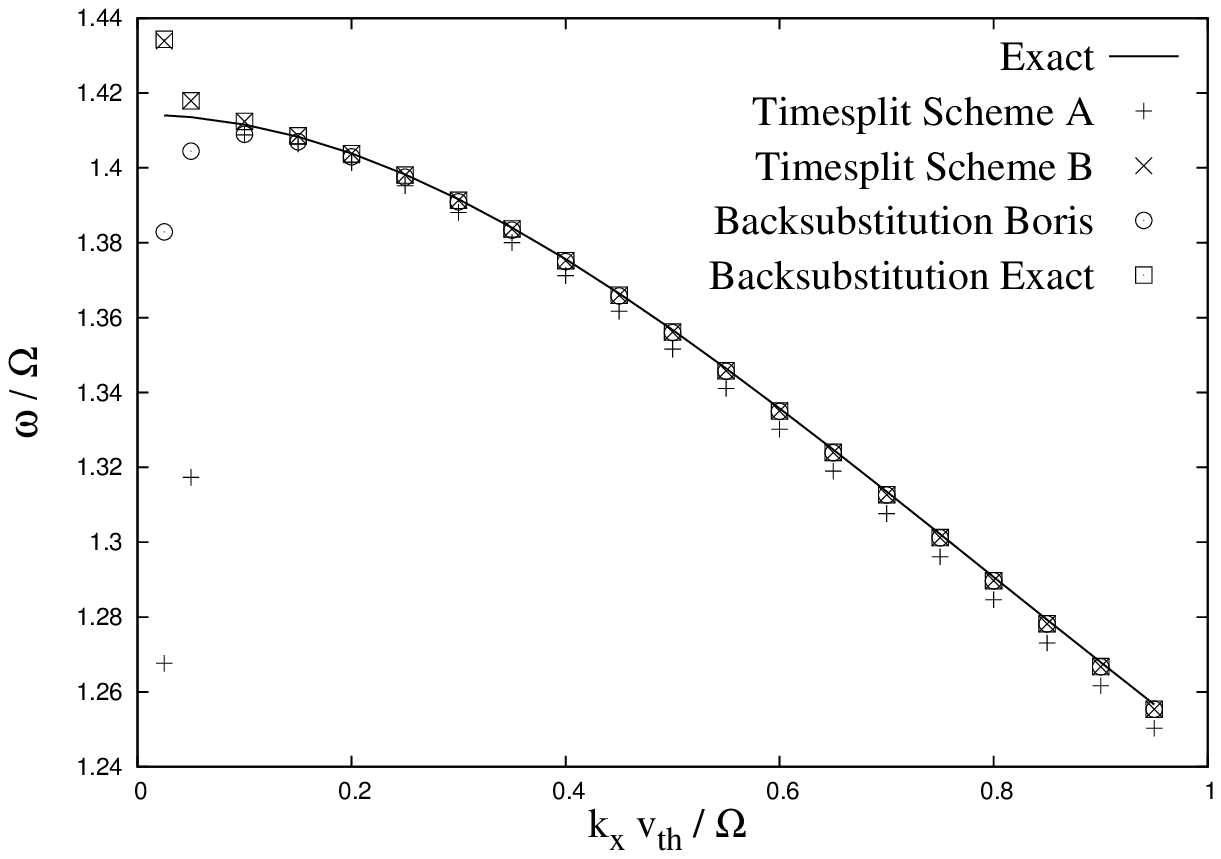}
\end{center}
\caption{Comparison of the dispersion of Bernstein waves between the exact
result and the Vlasov simulation using different schemes for integrating
the characteristics.
\label{FigDispersion}}
\end{figure}

The first simulation was initialised with the Bernstein wave of the lowest
frequency mode $1\le \omega/\Omega < 2$. Runs were performed for different
values of $k_x$ and using the different integration schemes. The frequency
of the wave was then determined using a Fourier analysis. The results are
shown in Fig. \ref{FigDispersion} for the time-splitting scheme A,
time-splitting scheme B, backsubstitution using the Boris scheme and
backsubstitution using the exact characteristics. We can observe that the
time-splitting scheme A clearly shows the largest error in the dispersion
of the waves. The errors of all the other schemes appear comparable and
are very good for all values of $k\ge0.15$. The larger errors for smaller
wavenumbers are due to the choice of the timestep $\Delta t$. Inserting
the definitions for $\Delta x$ and $k$ one finds that
\begin{equation}
\Delta t = 2\pi\Omega^{-1} \frac{1}{320 k}.
\end{equation}
For $k=0.05$ this means $\Delta t = 2\pi\Omega^{-1} / 16$ or 16 steps
for one gyration. 

\begin{figure}
\begin{center}
\includegraphics[width=10cm]{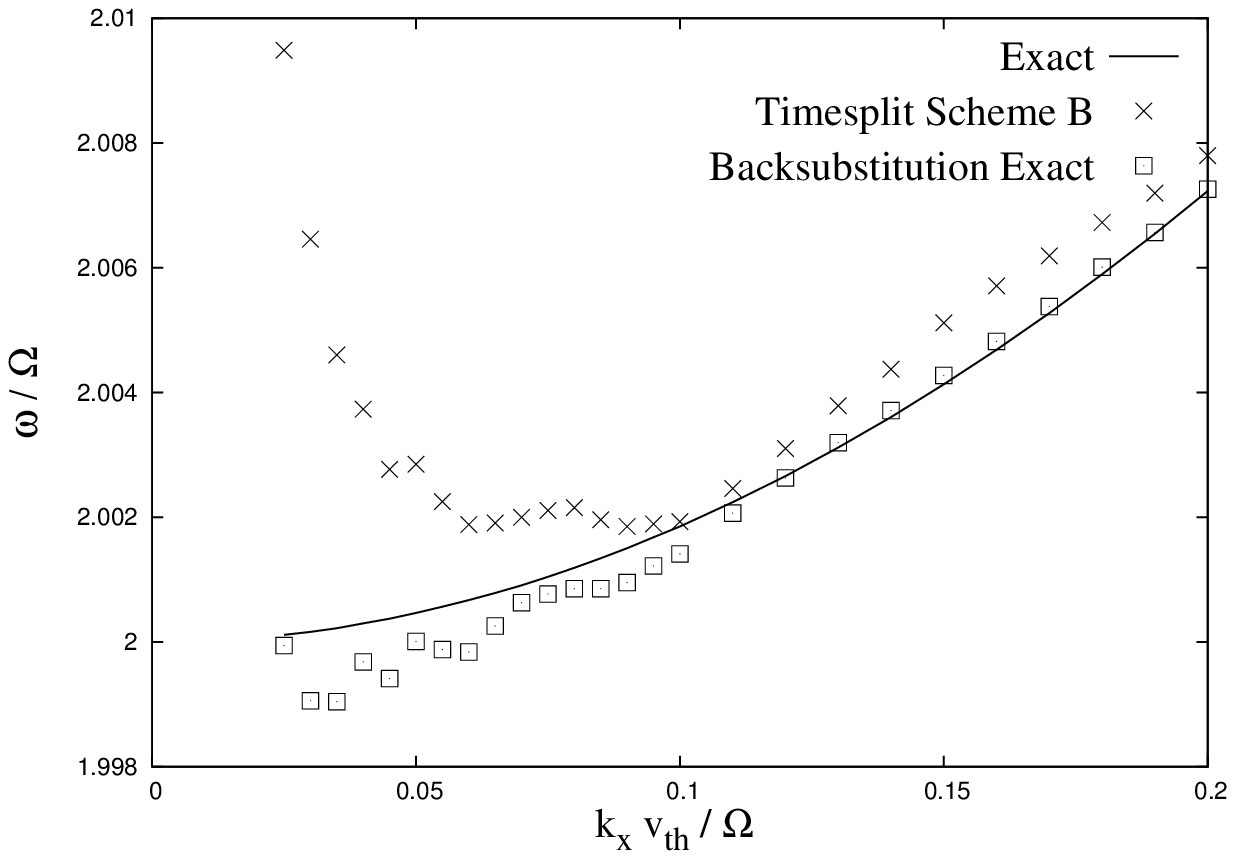}
\end{center}
\caption{Comparison of the dispersion of Bernstein waves between the exact
result and the Vlasov simulation using different schemes for integrating
the characteristics.
\label{FigDispersion2}}
\end{figure}

In another simulation run, the second Bernstein mode $2 \le \omega/\Omega <
3$ was initialised. Fig. \ref{FigDispersion2} shows the result of the
time-splitting scheme B and the backsubstitution method with exact
integration of the characteristics. The dispersion relation is shown for
values of $k\le 0.2$, where the errors are largest due to the choice of
the timestep. For this case we observe that the backsubstitution show
superior results when compared to the time-splitting method.

\begin{table}[t]
\begin{center}
\begin{tabular}{|c|c|}
\hline
{\bf Scheme }      & {\bf Comp. time / min} \\ \hline \hline
Splitting Scheme A &  192  \\ \hline
Splitting Scheme B &  165  \\ \hline
Backsubs.\ Boris   &  94   \\ \hline
Backsubs.\ Exact   &  137  \\ \hline
\end{tabular}
\caption{Computational time used for a typical run
\label{TabCompTime}}
\end{center}
\end{table}

Finally, we want to look at the computational time used by the different
schemes. Table \ref{TabCompTime} shows the times used for a typical run.
The runs for the different schemes were carried out with exactly the same
conditions on the same machine. Here we find a clear advantage of the
backsubstitution scheme over the splitting schemes. The backsubstitution
scheme with Boris integration of the characteristics reduces the
computational effort by more than 50\% when compared to the time splitting
scheme A. With time-splitting scheme B this improvement is still
approximately 43\%. The timing for the exact backsubstitution shows less
improvement due to the fact that trigonometric functions have to be
evaluated. The reason for the speed--up is the fact that the
backsubstitution method has to integrate the distribution function only
once for each velocity dimension $v_x$, $v_y$ and $v_z$. The splitting
schemes, on the other hand, have to integrate the distribution function 7
times. Although the numerical effort of integrating the characteristics in
each step is considerably smaller in the splitting scheme B, this is
only a part of the computational time spent. Other parts involve the
interpolation of the distribution function and the calculation of fluxes
across the cell boundaries. Considering all the above results, the
backsubstitution method together with the Boris scheme can be taken as a
good alternative to the traditional time splitting method if speed is the
major issue. To obtain the most accurate results, the backsubstitution
method together with the exact integration of the characteristics is the
superior scheme. In addition it also is slightly faster than the
time-splitting scheme.

\section{Conclusions}

We have shown that the time splitting method for integrating Vlasov's
equation in higher dimensions can introduce systematic errors when used in
the presence of a magnetic field. These errors originate from the
effective splitting of the integration of the characteristics, when a
higher order integration scheme is used. The errors cause the temperature
of the distribution function to increase over time, and thus artificially
feed energy into the system.

The backsubstitution method presented here for the general case of
arbitrary integration schemes of the characteristics eliminates this
problem. Here not those characteristics that pass through the grid point
are integrated, but those characteristics that will give a consistent
scheme when executed in sequence for the full timestep. This not only
provides the best accuracy possible but also reduces the number of
integration steps. While in three dimensional velocity space the
time-splitting scheme consists of 7 steps, the backsubstitution method
only uses 3 steps since each component needs to be integrated only once.
Due to this advantage the backsubstitution method together with the Boris
scheme typically decreases the computational effort by over 40\% as
compared to a simple time splitting method while the errors remain small.
On the other hand, highest accuracy can be achieved with the
backsubstitution method together with the exact integration of the
characteristics.

\section*{Acknowledgements}

Access to the JUMP multiprocessor computer at the FZ J\"ulich was made
available through project HBO20. This work was supported by the SFB
591 of the Deutsche Forschungsgesellschaft.

\bibliographystyle{elsart-num} 

\begin{thebibliography}{1}
\bibitem{CHNG76}
C.~Z. Cheng, G.~Knorr, 
The integration of the Vlasov equation in configuration space,
J. Comp.\ Phys. {\bf 22} (1976) 330.

\bibitem{ARA66}
A.~Arakawa, Computational Design for Long-Term Numerical Integration of the
Equations of Fluid Motion: Two dimensional Incompressible Flow, Part 1,
J. Comp.\ Phys. {\bf 1} (1966) 119. Reprinted in J. Comp.\ Phys. 135 (1997) 103

\bibitem{FIJ99}
E.~Fijalkow, A numerical solution to the Vlasov equation, Comput.\ Phys.\
Communications {\bf 116} (1999) 319.

\bibitem{SON98}
E.~Sonnendr\"ucker, J.~Roche, P.~Bertrand, a.~Ghizzo, The Semi-Lagrangian
Method for the Numerical Resolution of Vlasov Equations,
J. Comp.\ Phys. {\bf 149} (1998) 201

\bibitem{FIL03}
F.~Filbet, E.~Sonnendr\"ucker,
Comparison of Eulerian Vlasov Solvers,
Comput.\ Phys.\ Communications {\bf 150} (2003) 247.

\bibitem{CAL01}
F.~Califano, A.~Mangeney, C.~Cavazzoni, P.~Travnicek,
A numerical scheme for the integration of the Vlasov--Maxwell system 
of equations,
in: Science and Supercomputing at CINECA, 2001, p. 456.

\bibitem{FIL01}
F.~Filbet, E.~Sonnendr\"ucker, P.~Bertrand,
Conservative numerical schemes for the Vlasov equation,
J. Comp.\ Phys. {\bf 172} (2001) 166.

\bibitem{ARB02}
T.~Arber, R.~G.~L. Vann,
A critical comparison of Eulerian grid based Vlasov solvers,
J. Comp.\ Phys. {\bf 180} (2002) 339.

\bibitem{BIR85}
C.~K. Birdsall, A.~B. Langdon,
Plasma Physics via Computer Simulation,
McGraw-Hill, New York, 1985.

\bibitem{NICH83}
D.~R. Nicholson, Introduction to plasma theory, 
John Wiley \& Sons, New York, 1983.

\bibitem{BER58}
I.~B. Bernstein,
Waves in a Plasma in a Magnetic Field,
Phys.\ Rev. {\bf 109} (1958) 10.


\end{thebibliography}

\end{document}